\documentclass{aa}
\usepackage[varg]{txfonts}
\usepackage{graphicx}
\usepackage{amsmath}
\usepackage[colorlinks=true,linkcolor=blue,citecolor=blue,urlcolor=blue]{hyperref}
\usepackage[switch]{lineno}

\AtBeginDocument{%
  \setlength{\abovedisplayshortskip}{\abovedisplayskip}%
  \setlength{\belowdisplayshortskip}{\belowdisplayskip}%
}

\begin{document}

\title{Where Do Quasars Live? DESI DR1 Constraints from PAC Measurements}

\titlerunning{Where Do Quasars Live?}
\authorrunning{Gui et al.}

\author{Shanquan Gui\inst{1}\thanks{\email{guishq25@mpa-garching.mpg.de}}
\and Kun Xu\inst{2,3,4}
\and Donghai Zhao\inst{5,6}}

\institute{Max Planck Institute for Astrophysics, Karl-Schwarzschild-Str. 1, D-85741 Garching, Germany
\and Center for Particle Cosmology, Department of Physics and Astronomy, University of Pennsylvania, Philadelphia, PA 19104, USA
\and Institute for Computational Cosmology, Department of Physics, Durham University, South Road, Durham DH1 3LE, UK
\and State Key Laboratory of Dark Matter Physics, Tsung-Dao Lee Institute \& School of Physics and Astronomy, Shanghai Jiao Tong University, Shanghai 201210, P.R. China
\and College of Physics and Electronic Information Engineering, Guilin University of Technology, Guilin 541004, China
\and Key Laboratory of Low-dimensional Structural Physics and Application, Education Department of Guangxi Zhuang Autonomous Region, Guilin 541004, China}

\date{Received xxx; accepted xxx}

\abstract
{Quasar environments connect the growth of supermassive black holes, active galactic nucleus feedback, and galaxy evolution. Small-scale quasar clustering probes the one-halo regime and can test whether quasar activity depends on central--satellite status, but spectroscopic-survey measurements are limited by quasar rarity and fiber collisions.}
{We seek to obtain precise small-scale measurements of quasar environments at $0.8<z<1.0$ and use them to constrain the quasar--halo connection, particularly the relative probability for satellite subhalos and central halos of the same halo accretion mass to host a quasar.}
{We apply the Photometric Objects Around Cosmic Webs (PAC) method to DESI Data Release~1 quasars and photometric galaxies from the DESI Legacy Imaging Surveys DR9. By exploiting the large photometric galaxy sample, PAC enables measurements of the excess projected surface density of neighbouring galaxies around quasars, $\bar{n}_{2}w_{\rm p}$, down to a stellar mass of $M_{\ast}=10^{10.80}M_{\odot}$ over $0.1<r_{\rm p}/(h^{-1}\,\mathrm{Mpc})<15$. To constrain the stellar-to-halo mass relation and the quasar--halo connection, we jointly model the excess surface-density measurements, the quasar and luminous red galaxy (LRG) autocorrelation functions, and the quasar--LRG cross-correlation. We interpret these measurements using an N-body simulation together with a stellar-to-halo mass relation, an explicit stellar-mass-incompleteness model, and a Gaussian quasar occupation as a function of halo accretion mass.}
{The Gaussian quasar occupation peaks at $\log_{10}(M_{\rm acc}/h^{-1}M_{\odot})=12.88^{+0.02}_{-0.02}$ with width $\sigma_{\rm q}=0.51^{+0.02}_{-0.01}$, and the relative satellite-hosting parameter, $B$, defined as the quasar-hosting probability of a satellite subhalo relative to that of a central halo at fixed halo accretion mass, is $B=1.01^{+0.03}_{-0.03}$.}
{Within the adopted model framework, quasars are consistent with being equally likely to reside in central halos and satellite subhalos at fixed halo accretion mass. PAC has strong potential to deliver precise small-scale measurements of quasar environments.}


\maketitle

\makeatletter
\fancyhead[CE]{\aa@headfont \aa@headings}
\makeatother

\section{Introduction}\label{sec:introduction}

The formation and evolution of galaxies are closely connected to the growth of the supermassive black holes (SMBHs) at their centers.  Quasars represent a luminous phase of SMBH accretion and therefore provide an important probe of black-hole growth over cosmic time.  Observed scaling relations between SMBHs and their host galaxies, together with theoretical models of gas accretion and feedback, suggest that black-hole growth and galaxy evolution are coupled processes \citep{kauffmann2000,kormendy2013,heckman2014,terrazas2025}.  AGN feedback can heat, expel, or redistribute gas in galaxies and their surrounding halos, making the environments of quasars an important diagnostic of the physical conditions that trigger and sustain SMBH growth \citep{fabian2012,king2015,laha2021,harrison2024}.

Quasar environments provide a direct observational link between SMBH growth and the large-scale structure in which galaxies form. Different fueling channels, including galaxy mergers, tidal interactions, disk instabilities, and secular gas inflows, can depend on halo mass, local density, and halo assembly history \citep{hopkins2008}. Measurements of quasar clustering and quasar--galaxy correlations therefore constrain the halo environments in which different triggering channels operate, while not by themselves identifying a unique triggering mechanism.

In the standard picture of hierarchical structure formation, the large-scale environments of galaxies and quasars are set primarily by their host dark-matter halos.  Measurements of large-scale quasar clustering have shown that quasars typically occupy halos with masses of a few times $10^{12}\,h^{-1}M_{\odot}$ over a broad redshift range \citep{croom2005,ross2009,shen2009,white2012,eftekharzadeh2015,laurent2017,hou2021}.  These measurements robustly constrain the mean host-halo mass and bias of quasars, but they do not by themselves determine how quasars populate halos on small scales.

Small-scale clustering is particularly important because it probes the one-halo regime and is sensitive to the relative occupation of central halos and satellite subhalos. If quasar activity is enhanced by interactions, mergers, or other environmental processes, then subhalos may have a different probability of hosting a quasar than central halos of the same halo accretion mass. Conversely, if quasar activity is controlled mainly by halo mass or internal galaxy properties, the occupation probability may depend only weakly on central--satellite status. The relative quasar-hosting probability at fixed halo accretion mass is therefore a useful, model-dependent diagnostic of the environmental dependence of SMBH growth \citep{richardson2012,chatterjee2013,kayo2012,chowdhary2025qsohod}.

Existing observational constraints on small-scale quasar occupation remain uncertain. Quasars are rare, so small-scale measurements in spectroscopic surveys suffer from substantial shot noise. Fiber collisions can further reduce the sampling of close angular pairs, limiting the precision of small-scale clustering measurements \citep{hennawi2006,myers2007,richardson2012}. Quasar halo-occupation studies have therefore obtained useful, but model-dependent, constraints on satellite occupation and duty cycle \citep{chatterjee2013,shen2013,eftekharzadeh2019}.

The Photometric Objects Around Cosmic Webs (PAC) method addresses these limitations by combining spectroscopic tracers with photometric galaxy samples \citep{xu2022b,xu2023}. PAC measures the excess projected surface density of photometric galaxies around spectroscopic targets, thereby avoiding reliance on close spectroscopic pairs. The high surface density of the photometric sample improves the precision of environmental measurements and retains stellar-mass information for the neighbouring galaxies. Previous PAC analyses have shown that this approach recovers both one-halo and two-halo information and constrains the galaxy--halo connection \citep{gui2024}.

In this work, we apply PAC to DESI Data Release~1 quasars and photometric galaxies from the DESI Legacy Imaging Surveys DR9. We focus on quasars at $0.8<z<1.0$ and measure the excess projected galaxy surface density, $\bar{n}_{2}w_{\rm p}$, over $0.1<r_{\rm p}/(h^{-1}\,\mathrm{Mpc})<15$. We interpret the measurements using an N-body simulation together with a stellar-to-halo mass relation model and a quasar--halo connection model. The parameter $B$ is defined as the quasar-hosting probability of a subhalo relative to that of a central halo with the same halo accretion mass; it is not itself the quasar satellite fraction. Our fiducial analysis gives $B=1.01^{+0.03}_{-0.03}$, consistent with equal probabilities at fixed halo accretion mass within the adopted model.

This paper is organized as follows.  Section~\ref{sec:methods} describes the observational samples and PAC measurements.  Section~\ref{sec:sim_results} presents the simulation, SHAM model, magnification treatment, and quasar--halo connection constraints.  Section~\ref{sec:discussion} discusses the implications and limitations of the results.  Section~\ref{sec:conclusions} summarizes our conclusions.

\section{Data and measurements}\label{sec:methods}

We first describe the DESI spectroscopic tracers and the Legacy Surveys photometric catalogue. We then define the imaging-depth cuts, stellar-mass estimates, and completeness treatment for the photometric sample. Finally, we present the PAC estimator, covariance calculation, redshift-bin combination, and additional spectroscopic clustering measurements used in the model fits.

\subsection{Spectroscopic and photometric sample}\label{subsec:survey}

Following the PAC framework \citep{gui2024}, which combines spectroscopic and photometric surveys, we use wide-area optical imaging from Data Release 9 of the DESI Legacy Imaging Surveys as the parent photometric catalogue \citep{dey2019} and DESI DR1 spectroscopic quasar data as the spectroscopic sample. The Legacy Surveys DR9 catalogues\footnote{\url{https://www.legacysurvey.org/dr9/catalogs/}} and the DESI DR1 LSS catalogues used for the spectroscopic sample\footnote{\url{https://data.desi.lbl.gov/public/dr1/survey/catalogs/dr1/LSS/iron/LSScats/v1.5/}} are publicly available.

The DECaLS imaging covers approximately $9000\,\mathrm{deg}^{2}$ in the Northern and Southern Galactic Caps at declination $<32^{\circ}$, and incorporates Dark Energy Survey imaging over an additional $\simeq5000\,\mathrm{deg}^{2}$ in the South Galactic Cap \citep{dey2019}.  The catalogue provides $grz$ photometry, with typical median $5\sigma$ point-source depths of $g=24.9$, $r=24.2$, and $z=23.3$.  Sources are extracted with \textsc{Tractor} \citep{lang2016}, which fits parametric surface-brightness models convolved with the local point-spread function.  We use the extinction-corrected model magnitudes, where the Galactic extinction correction follows \citet{schlegel1998}.  We restrict the analysis to regions observed at least once in all three bands and apply the bright-star and bad-pixel masks encoded in the Legacy Surveys \texttt{MASKBITS}; the bit definitions are documented online.\footnote{\url{https://www.legacysurvey.org/dr9/bitmasks/}}

The spectroscopic sample is drawn from the first year of DESI observations released in DESI DR1. DESI is a Stage-IV dark-energy experiment designed to obtain spectra for tens of millions of extragalactic targets with a 5000-fiber multi-object spectrograph on the 4 m Mayall Telescope \citep{levi2013,desi2016a,desi2016b,desi2022,desi2024a,desi2024b}. The instrument covers approximately $3600$--$9800$~\AA and can observe up to 5000 targets per pointing \citep{desi2016b,silber2023,miller2024,poppett2024}.

We select spectroscopically confirmed DESI quasars in the redshift interval $0.8<z_{\rm s}<1.0$. The DESI quasar target selection combines optical and infrared photometry with variability information and is designed to extend to $z>2$ \citep{DESIQSO}.  After these cuts, the quasar sample contains 72,855 objects.

To constrain the stellar--halo mass relation, we also include DESI spectroscopic luminous red galaxies (LRGs) in the same redshift interval, $0.8<z_{\rm s}<1.0$.  DESI LRG targets are selected from Legacy Surveys $grz$ photometry together with WISE $W1$ information using color--magnitude cuts designed to isolate old stellar populations while controlling stellar contamination and imaging systematics \citep{zhou2023lrg}.

\subsection{Photometric depth and stellar-mass estimates}\label{subsec:sed}

To minimize spatially varying imaging-depth effects, we restrict both the photometric and spectroscopic samples to the DECaLS footprint used for the PAC measurements.

\begin{figure}[!t]
    \centering
    \includegraphics[width=0.9\linewidth]{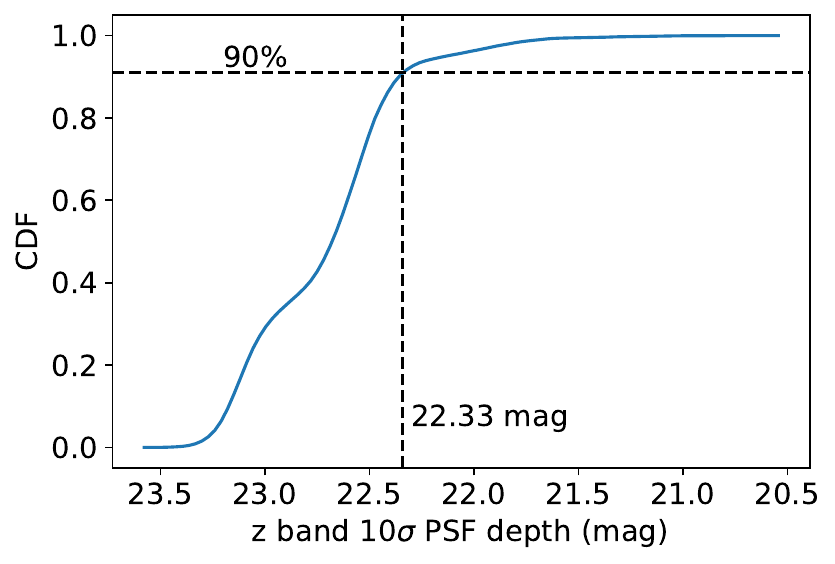}
    \caption{Cumulative area distribution of the DECaLS $z$-band $10\sigma$ point-source depth in the DESI Legacy Imaging Surveys. The vertical reference depth corresponds to the depth reached by approximately 90\% of the DECaLS footprint.}
    \label{fig:decalse_zban}
\end{figure}

Following the PAC analysis \citep{gui2024}, we characterize the imaging depth with the $z$-band $10\sigma$ point-source depth. Fig.~\ref{fig:decalse_zban} shows the cumulative area distribution of the DECaLS $z$-band depth. Approximately 90\% of the DECaLS area reaches depths fainter than $z=22.33$. We therefore adopt this value as the representative depth limit for defining the uniform photometric sample.

The stellar masses of photometric objects and LRGs are estimated with the SED-fitting code CIGALE \citep{boquien2019} using the observed $grz$ fluxes.  The stellar population models are based on the Bruzual--Charlot synthesis library and assume a Chabrier initial mass function \citep{bruzual2003,chabrier2003}.  We allow three stellar metallicities, $Z/Z_{\odot}=0.4$, 1.0, and 2.5.  The star-formation history is parameterized as a delayed model, $\mathrm{SFR}(t) \propto t\exp(-t/\tau)$, where the timescale $\tau$ spans $10^{7}$--$1.258\times10^{10}\,\mathrm{yr}$ in logarithmic steps of 0.1 dex.  Dust attenuation is modeled with the Calzetti starburst attenuation law \citep{calzetti2000}, with the color excess $E(B-V)$ varied between 0 and 0.5.

\begin{figure}[!t]
    \centering
    \includegraphics[width=1.0\linewidth]{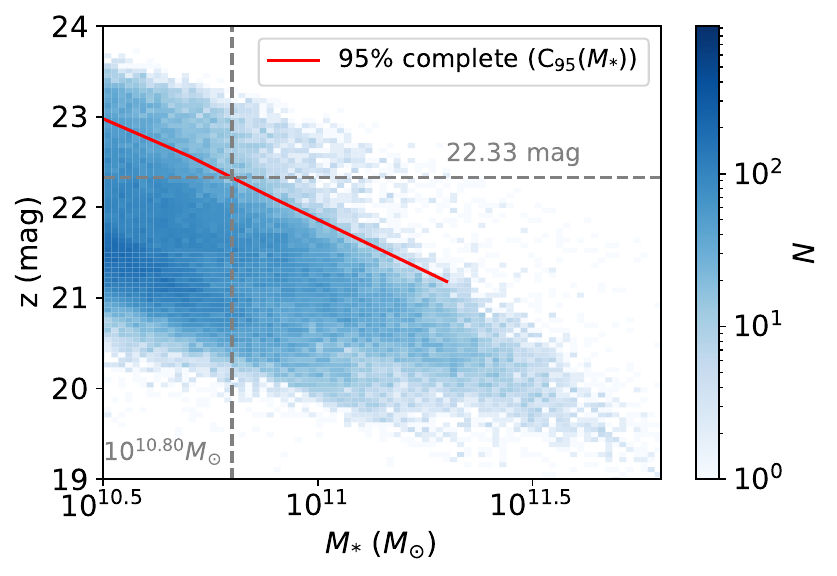}
    \caption{Stellar-mass completeness based on the deepest Legacy Surveys regions. The red curve marks $C_{95}(M_{\ast})$, defined as the $z$-band magnitude brighter than which 95\% of galaxies at fixed stellar mass are retained.}
    \label{fig:decalse_completeness}
\end{figure}

We estimate stellar-mass completeness. A deep reference sample is constructed from the deepest $50\,\mathrm{deg}^{2}$ of the Legacy Surveys, corresponding to 738 bricks with $z$-band $10\sigma$ point-source depth deeper than 23.37. Photometric redshifts for this reference sample are taken from the DESI Legacy Surveys photo-$z$ catalogue \citep{zhou2021}. For each stellar-mass bin, we define $C_{95}(M_{\ast})$ as the 95th percentile of the $z$-band magnitude distribution. Fig.~\ref{fig:decalse_completeness} illustrates this stellar-mass completeness identification for $0.8<z_{\rm p}<1.0$.

At the adopted DECaLS depth limit of $z=22.33$, the stellar-mass completeness threshold reaches
$\log(M_{\ast}/M_{\odot})=10.80$ at $0.8<z<1.0$.
Therefore, we use photometric galaxies with stellar masses in the range
$10^{10.80}\leq M_{\ast}/M_{\odot}\leq10^{11.80}$ as the complete sample,
divided into four equal stellar-mass bins.
Furthermore, to constrain the SHMR at lower masses, we incorporate the incomplete stellar-mass bins within the range
$10^{10.00}\leq M_{\ast}/M_{\odot}<10^{10.80}$
through explicit incompleteness modelling, as described below.

\subsection{Measurements}\label{subsec:measurements}

We measure the excess projected surface density of photometric galaxies around spectroscopic quasars using the PAC formalism \citep{xu2022b}. The measured quantity is
\begin{equation}
    A(r_{\rm p}) \equiv \bar{n}_{2}w_{\rm p}(r_{\rm p}),
\end{equation}
where $\bar{n}_{2}$ is the mean comoving number density of the photometric galaxy sample and $r_{\rm p}$ is the projected comoving separation.

We compute the signal with a Landy--Szalay-type estimator \citep{landy1993} in four redshift bins, $0.80<z_{\rm s}<0.85$, $0.85<z_{\rm s}<0.90$, $0.90<z_{\rm s}<0.95$, and $0.95<z_{\rm s}<1.00$. The PAC signal is measured independently in each redshift bin.
For the $i$th redshift bin, we estimate the mean PAC signal from
$N_{\rm sub}$ jackknife subsamples as

\begin{equation}
A_i^a
=
\frac{1}{N_{\rm sub}}
\sum_{k=1}^{N_{\rm sub}}
A_{i,k}^a ,
\end{equation}

where $A_{i,k}^a$ denotes the PAC measurement in the
$a$th radial bin from the $k$th jackknife realization.

The covariance matrix for the $i$th redshift bin is estimated
using the standard jackknife estimator,

\begin{equation}
C_i^{ab}
=
\frac{N_{\rm sub}-1}{N_{\rm sub}}
\sum_{k=1}^{N_{\rm sub}}
\left(
A_{i,k}^{a}-A_i^{a}
\right)
\left(
A_{i,k}^{b}-A_i^{b}
\right),
\label{eq:cov_redshift_bin}
\end{equation}

where $a$ and $b$ denote the $a$th and $b$th radial bins,
respectively.

The measurements from different redshift bins are then combined
using inverse-variance weighting. Defining

\begin{equation}
W_i^a=\frac{1}{(\sigma_i^a)^2},
\end{equation}

where $(\sigma_i^a)^2=C_i^{aa}$ is the diagonal element of the
covariance matrix, the combined PAC signal is

\begin{equation}
A^a
=
\frac{
\sum_{i=1}^{N_z}
W_i^a A_i^a
}{
\sum_{i=1}^{N_z}
W_i^a
},
\label{eq:combined_signal}
\end{equation}

where $N_z$ is the number of redshift bins. The covariance matrix of the combined
signal is

\begin{equation}
C^{ab}
=
\frac{
\sum_{i=1}^{N_z}
W_i^a W_i^b
C_i^{ab}
}{
\left[
\sum_{i=1}^{N_z}W_i^a
\right]
\left[
\sum_{i=1}^{N_z}W_i^b
\right]
}.
\label{eq:combined_covariance}
\end{equation}

\begin{figure*}[!t]
    \centering
    \includegraphics[width=0.8\linewidth]{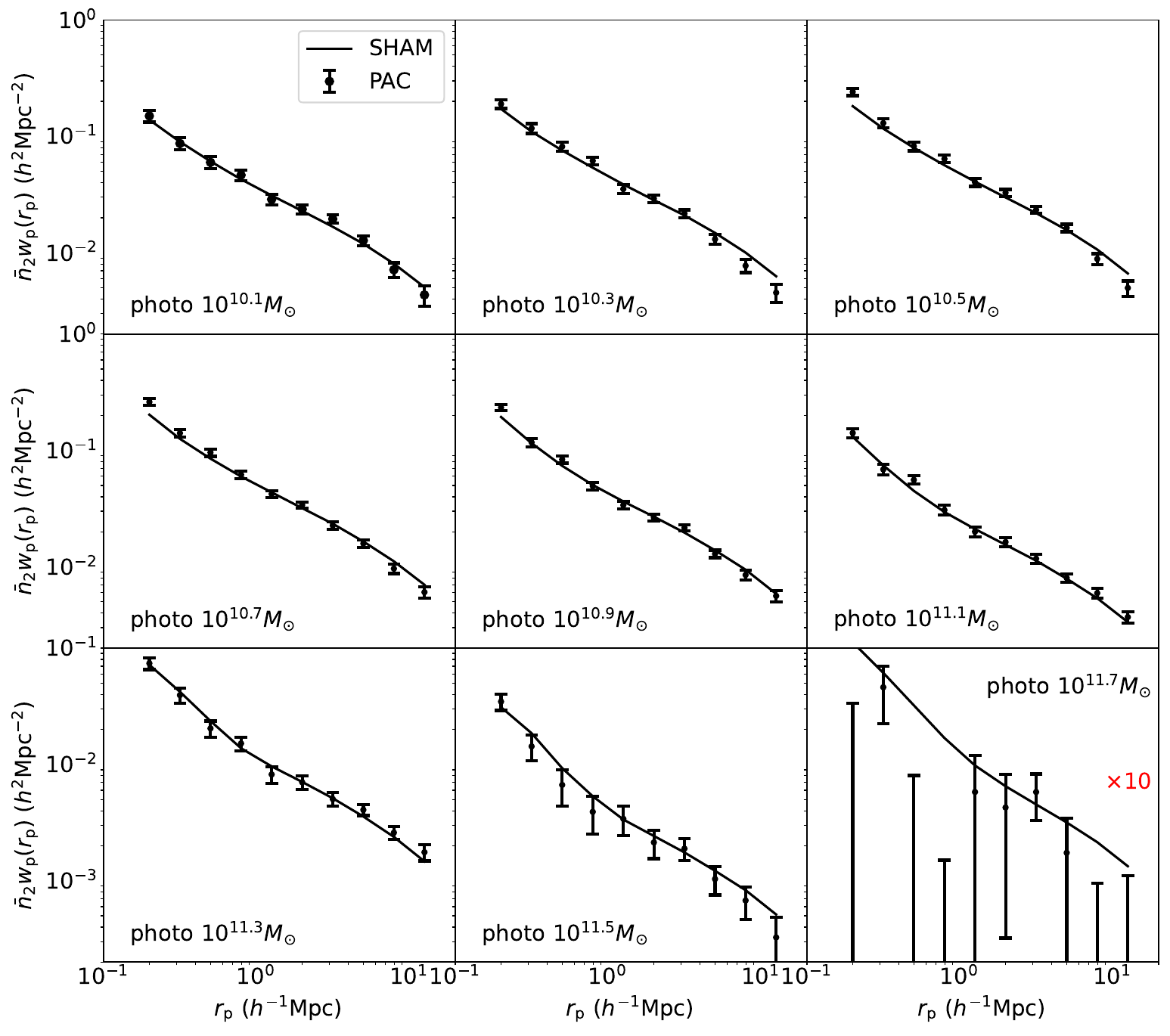}
    \caption{Measurements (points with error bars) and best-fitting model (lines) for the excess projected density $\bar{n}_{2}w_{\rm p}$ of photometric galaxies around DESI DR1 quasars. The galaxy stellar-mass range is $10^{10.0}\leq M_{\ast}/M_{\odot}\leq10^{11.8}$, with the median stellar mass indicated in each panel. The highest-stellar-mass bin is multiplied by 10 for visual clarity.}
    \label{fig:withmag_pre}
\end{figure*}

\begin{figure}[!t]
    \centering
    \includegraphics[width=1.0\linewidth]{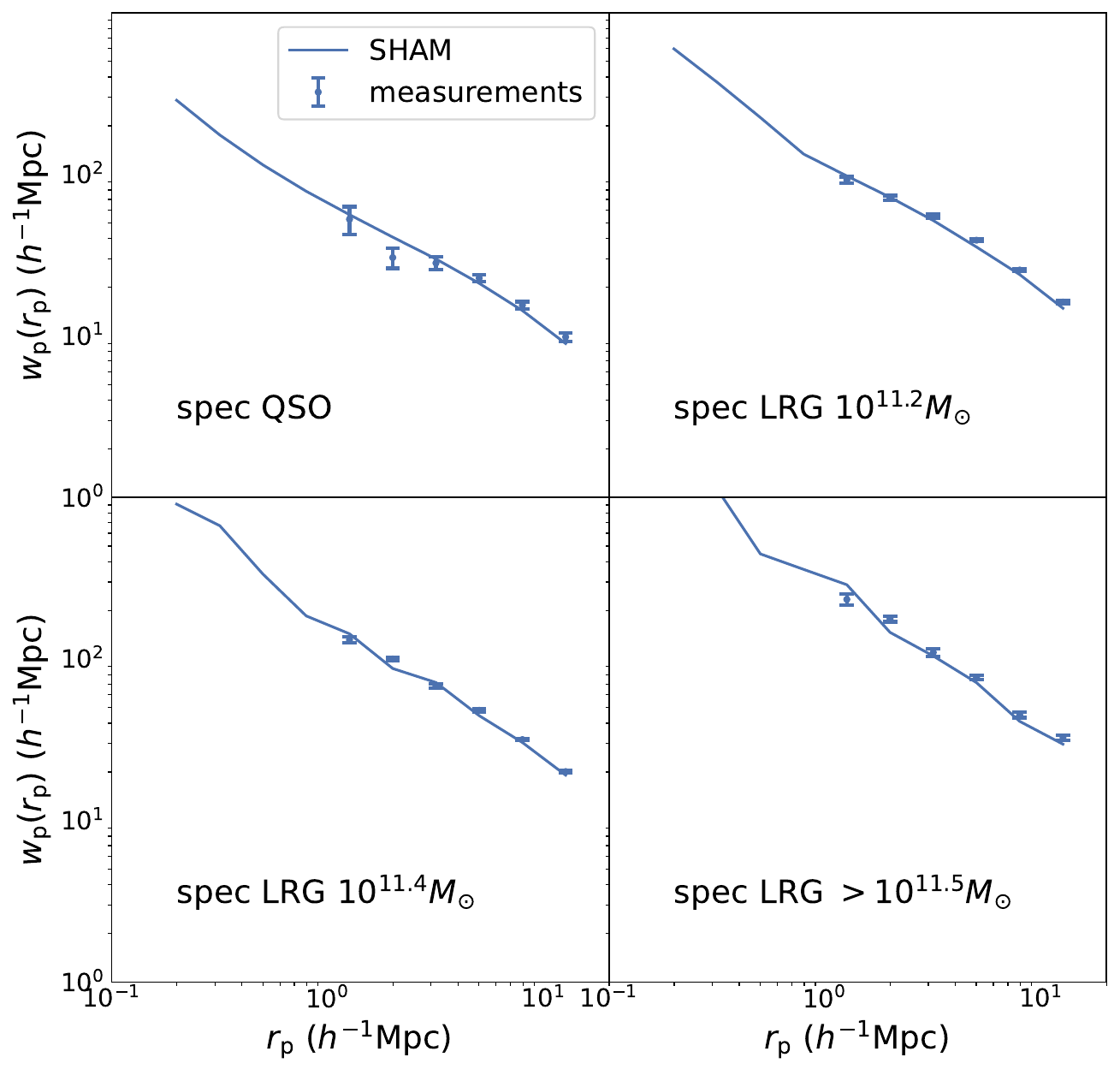}
    \caption{Measurements (points with error bars) and best-fitting model (lines) for the projected autocorrelation functions $w_{\rm p}$ of DESI DR1 quasars and LRGs.}
    \label{fig:auto_pre}
\end{figure}

\begin{figure*}[!t]
    \centering
    \includegraphics[width=1.0\linewidth]{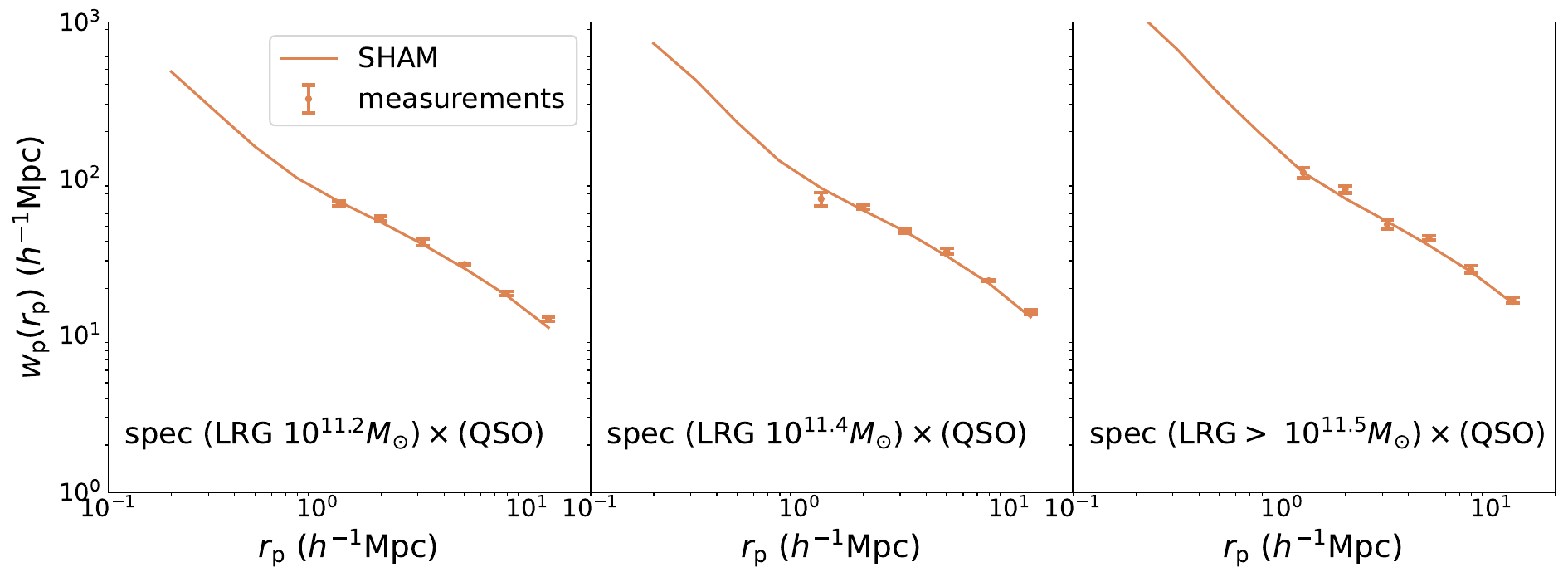}
    \caption{Measurements (points with error bars) and best-fitting model (lines) for the projected quasar--LRG cross-correlation function $w_{\rm p}$ of DESI DR1 quasars and LRGs.}
    \label{fig:cross_pre}
\end{figure*}

We measure $\bar{n}_{2}w_{\rm p}(r_{\rm p})$ over $0.1<r_{\rm p}/(h^{-1}\,\mathrm{Mpc})<15$, covering both the one-halo and two-halo regimes. We also include the projected autocorrelation functions of quasars and LRGs and the quasar--LRG projected cross-correlation to constrain the stellar--halo mass relation and quasar--halo connection jointly. The LRGs are divided into three stellar-mass bins, $11.1\leq\log(M_{\ast}/M_{\odot})<11.3$, $11.3\leq\log(M_{\ast}/M_{\odot})<11.5$, and $\log(M_{\ast}/M_{\odot})\geq11.5$. For the spectroscopic correlation functions, we conservatively restrict the fit to $r_{\rm p}\geq1\,h^{-1}\mathrm{Mpc}$, where fiber-collision effects are negligible. Each object is assigned the total clustering weight
\begin{equation}
    w_{\rm tot}=w_{\rm comp}w_{\rm zfail},
\end{equation}
where $w_{\rm comp}$ corrects for incompleteness from fiber collisions and $w_{\rm zfail}$ accounts for variations in the redshift-success rate \citep{ross2025lss}. The PAC measurements are shown in Fig.~\ref{fig:withmag_pre}; the quasar/LRG autocorrelations and quasar--LRG cross-correlation are shown in Figs.~\ref{fig:auto_pre} and \ref{fig:cross_pre}.

\section{Modelling framework and results}\label{sec:sim_results}

In this section, we describe the N-body simulation, the SHAM model, and the quasar--halo assignment used to interpret the PAC measurements. We also quantify the foreground-magnification systematic in the PAC signal and assess its impact on the inferred quasar--halo parameters.

\subsection{N-body simulation}\label{subsec:nbody}

We employ the CosmicGrowth simulation suite \citep{jing2019}, a set of high-resolution N-body simulations run with the adaptive parallel $\mathrm{P}^{3}\mathrm{M}$ method \citep{jing2002,xu2021}.  We use the flat $\Lambda$CDM run with cosmological parameters $\Omega_{\rm m}=0.268$, $\Omega_{\Lambda}=0.732$, $h=0.71$, $n_{\rm s}=0.968$, and $\sigma_{8}=0.83$.  The simulation contains $3072^{3}$ dark-matter particles in a cubic box of side length $600\,h^{-1}\mathrm{Mpc}$, with a particle mass $m_{\rm p}=5.54\times10^{8}\,h^{-1}M_{\odot}$ and a force softening length $\eta=0.01\,h^{-1}\mathrm{Mpc}$.  Dark-matter halos are identified with the friends-of-friends algorithm using a linking length of 0.2 times the mean particle separation \citep{davis1985}.  Subhalos and their merger histories are constructed with \textsc{HBT+} \citep{han2012,han2018}; for subhalos falling below 20 particles, merger timescales are estimated using the fitting formula of \citet{jiang2008}.  We use snapshot 76, at $z\simeq0.92$, to match the effective redshift of the quasar and LRG samples.

\subsection{Galaxy--halo and quasar--halo models}\label{subsec:sham_model}

We connect galaxies to halos and subhalos through subhalo abundance matching. We describe the stellar-to-halo mass relation with a five-parameter double power law and a constant log-normal scatter \citep{wang2010,yang2012,moster2013,behroozi2019,xu2023}:
\begin{equation}\label{eq:shmr}
    M_{\ast} =
    \frac{2k}{\left(M_{\rm acc}/M_{0}\right)^{-\alpha}
    + \left(M_{\rm acc}/M_{0}\right)^{-\beta}} .
\end{equation}
Here $M_{\rm acc}$ is the virial mass of a halo or subhalo at the last epoch when it was a central object, and $M_{\rm vir}$ follows the spherical-overdensity definition of \citet{bryan1998}. At fixed $M_{\rm acc}$, the scatter in $\log M_{\ast}$ is modelled as a Gaussian with width $\sigma$. The same SHMR parameters are applied to central and satellite galaxies. We allow the observed number density in the four lowest stellar-mass bins to be incomplete. The nuisance parameters $k_1$, $k_2$, $k_3$, and $k_4$ rescale the corresponding model predictions and are therefore effective selection factors rather than direct measurements of a physical missing fraction.

For quasars, we assume that the probability for a halo to host a quasar depends on accretion mass through a Gaussian in $\log M_{\rm acc}$,
\begin{equation}\label{eq:qso_probability}
    P_{\rm cen}(M_{\rm acc}) \propto
    \exp\left\{-\frac{\left[\log_{10}(M_{\rm acc}/h^{-1}M_{\odot})-\mu\right]^2}
    {2\sigma_{\rm q}^{2}}\right\} .
\end{equation}
The same $\mu$ and $\sigma_{\rm q}$ are used for halos and subhalos.  We allow the overall probability for a selected subhalo to differ from that of a selected central halo by a factor $B$.  For a fixed Gaussian selection, the quasar satellite fraction is therefore
\begin{equation}\label{eq:fsat}
    f_{\rm sat} = \frac{B N_{\rm sat}}{B N_{\rm sat}+N_{\rm cen}},
\end{equation}
where $N_{\rm sat}$ and $N_{\rm cen}$ are the numbers of selected subhalos and central halos before applying the factor $B$.  Thus $B$ characterizes the relative quasar-hosting probability of subhalos and central halos at fixed halo accretion mass: $B=1$ corresponds to equal probabilities, $B>1$ to enhanced satellite occupation, and $B<1$ to suppressed satellite occupation.

After assigning galaxies and quasars to simulated halos, we compute the projected auto- and cross-correlation functions with \textsc{Corrfunc} \citep{sinha2020}. The full model comparison is written in vector form as
\begin{equation}\label{eq:chisq_full}
    \chi^{2} = \left(\mathbf{d}-\mathbf{m}\right)^{\rm T}
    \mathbf{C}^{-1}
    \left(\mathbf{d}-\mathbf{m}\right),
\end{equation}
where $\mathbf{d}$ is the data vector containing the PAC measurements $\bar{n}_{2}w_{\rm p}$, the quasar and LRG autocorrelation functions $w_{\rm p}$, and the quasar--LRG cross-correlation function $w_{\rm p}$.  The vector $\mathbf{m}$ is the corresponding model prediction, and $\mathbf{C}$ is the measurement covariance matrix.  We explore the parameter space with the MCMC sampler \textsc{emcee} \citep{foremanmackey2013}, fitting the SHMR parameters $\{M_{0},\alpha,\beta,k,\sigma\}$, the incompleteness parameters $\{k_1,k_2,k_3,k_4\}$, and the quasar--halo parameters $\{\mu,\sigma_{\rm q},B\}$.

\subsection{Magnification systematics}\label{subsec:magnification}

\begin{figure}[!t]
    \centering
    \includegraphics[width=1.0\linewidth]{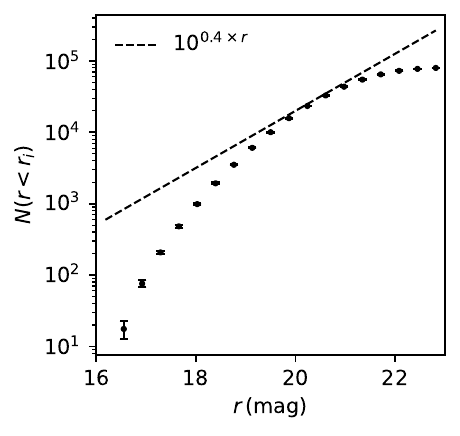}
    \caption{Cumulative DESI DR1 quasar number counts as a function of $r$-band magnitude. The dashed line marks the critical slope $s=0.4$.}
    \label{fig:slope}
\end{figure}

Foreground large-scale structure can weakly lens the DESI quasar sample and produce an apparent quasar--galaxy correlation even when the foreground galaxies are not physically associated with the quasars.  For a flux-limited background sample, magnification changes both the observed solid angle and the number of sources promoted across the selection limit.  If the cumulative number counts are locally described by
\begin{equation}
    s \equiv \frac{\mathrm{d}\log_{10}N_{\rm q}(<m)}{\mathrm{d}m},
\end{equation}
where $m$ denotes the apparent magnitude of the background quasar sample in the band used to define the number-count slope, and $N_{\rm q}(<m)$ is the cumulative number of quasars brighter than that magnitude limit.
The fractional change in the observed quasar surface density is
\begin{equation}\label{eq:magnification_density}
    \frac{\delta n_{\rm q}^{\rm mag}}{n_{\rm q}}
    = \mu^{2.5s-1}-1
    \simeq \left(5s-2\right)\kappa,
\end{equation}
where $\mu$ is the lensing magnification, $\kappa$ is the convergence, and the final expression uses the weak-lensing limit $\mu\simeq1+2\kappa$ \citep{moessner1998,bartelmann2001,scranton2005,kayo2012}. The magnification contribution changes sign at $s=0.4$. Fig.~\ref{fig:slope} shows the cumulative $r$-band number counts of the DESI DR1 quasar sample and the corresponding sign-change threshold. Because DESI quasar target selection depends on optical and infrared photometry \citep{DESIQSO}, we estimate an effective count slope for the multiband selection rather than using a single-band power-law approximation. Following multiband magnification-bias treatments \citep{myers2003,wyithe2003}, we define $s_{\rm eff}$ through
\begin{equation}\label{eq:effective_slope}
    \mu^{2.5s_{\rm eff}-1}
    =
    \frac{1}{\mu}
    \frac{N_{\rm q}\left(<\mathbf{m}_{\rm lim}+\Delta m_{\mu}\right)}
    {N_{\rm q}\left(<\mathbf{m}_{\rm lim}\right)} ,
\end{equation}
where $N_{\rm q}(<\mathbf{m})$ denotes the number of quasars passing all magnitude limits in the selection vector $\mathbf{m}$, $\Delta m_{\mu}=2.5\log_{10}\mu$, and $\mathbf{m}_{\rm lim}=(m_{r,{\rm lim}},m_{W1,{\rm lim}},m_{W2,{\rm lim}})$.  The factor $\mu^{-1}$ accounts for the dilution of the observed solid angle, while $\Delta m_{\mu}$ accounts for the flux amplification.  We use $m_{r,{\rm lim}}=23$ and $m_{W1,{\rm lim}}=m_{W2,{\rm lim}}=22.3$, matching the limits used for the DESI quasar selection in this analysis.  In the following calculation, $s$ in Equation~\ref{eq:magnification_density} is replaced by this multiband effective slope $s_{\rm eff}$.

We predict the magnification-induced signal from the simulation and convert it to the same $\bar{n}_{2}w_{\rm p}$ observable used for the PAC measurements.  We first compute the magnification excess for a single lens--source configuration.  For foreground galaxies of stellar mass $M_{\ast}$ at redshift $z_{\rm l}$ lensing a background quasar at redshift $z_{\rm s}$, the weak-lensing magnification excess is
\begin{equation}\label{eq:dmu_single}
    \delta\mu(\theta;M_{\ast},z_{\rm l},z_{\rm s})
    = \frac{2\Sigma(\theta;M_{\ast},z_{\rm l})}
    {\Sigma_{\rm crit}(z_{\rm l},z_{\rm s})} .
\end{equation}
Here $\Sigma(\theta;M_{\ast},z_{\rm l})$ is the projected surface density measured around the simulated foreground galaxies in the corresponding stellar-mass and redshift bin.  The critical surface density is
\begin{equation}\label{eq:sigma_crit}
    \Sigma_{\rm crit}(z_{\rm l},z_{\rm s})
    = \frac{c^2}{4\pi G}
    \frac{D_{\rm s}}{D_{\rm l}D_{\rm ls}},
\end{equation}
with the usual factor of $(1+z_{\rm l})^{-2}$ included when using comoving projected densities.

We then average the single-configuration magnification excess over the foreground galaxy distribution and the background quasar redshift distribution.  For foreground galaxies within the stellar-mass interval $M_{\ast,l}<M_{\ast}<M_{\ast,u}$, the mean magnification excess is
\begin{equation}\label{eq:dmu_all}
\begin{split}
    \delta\mu_{\rm all}(\theta)
    ={}&
    \int_{M_{\ast,l}}^{M_{\ast,u}}
    \int
    \int
    p_{\rm fg}(M_{\ast},z_{\rm l})\,
    p_{\rm bg}(z_{\rm s})
    \\
    &\times
    \delta\mu(\theta;M_{\ast},z_{\rm l},z_{\rm s})\,
    \mathrm{d}M_{\ast}\,
    \mathrm{d}z_{\rm l}\,
    \mathrm{d}z_{\rm s} .
\end{split}
\end{equation}
where $p_{\rm fg}(M_{\ast},z_{\rm l})$ is the normalized joint stellar-mass--redshift distribution of the foreground galaxies in the selected stellar-mass bin, measured from the photometric-redshift catalogue, and $p_{\rm bg}(z_{\rm s})$ is the normalized redshift distribution of the background quasars.

The angular quasar--foreground correlation generated by magnification is then
\begin{equation}\label{eq:magnification_wtheta}
    w_{\rm mag}(\theta)
    = \left\langle \delta_{\rm g}(\hat{\mathbf{n}})
    \delta_{\rm q}^{\rm mag}(\hat{\mathbf{n}}+\vec{\theta})\right\rangle
    = \left(2.5s_{\rm eff}-1\right)\delta\mu_{\rm all}(\theta).
\end{equation}
We evaluate this term for each foreground stellar-mass bin and apply the same angular-to-comoving binning as in the PAC measurement. The observed projected correlation function can be written as
\begin{equation}\label{eq:wp_obs_mag}
w_{\rm p}^{\rm obs}(r_{\rm p})
=
w_{\rm p}^{\rm phys}(r_{\rm p})
+
w_{\rm p}^{\rm mag}(r_{\rm p}) .
\end{equation}
Here $w_{\rm p}^{\rm phys}$ denotes the physical quasar--galaxy clustering signal modeled with the SHAM catalogue, while $w_{\rm p}^{\rm mag}$ denotes the projected contribution obtained from the angular magnification term $w_{\rm mag}(\theta)$.

\begin{figure}[!t]
    \centering
    \includegraphics[width=0.9\linewidth]{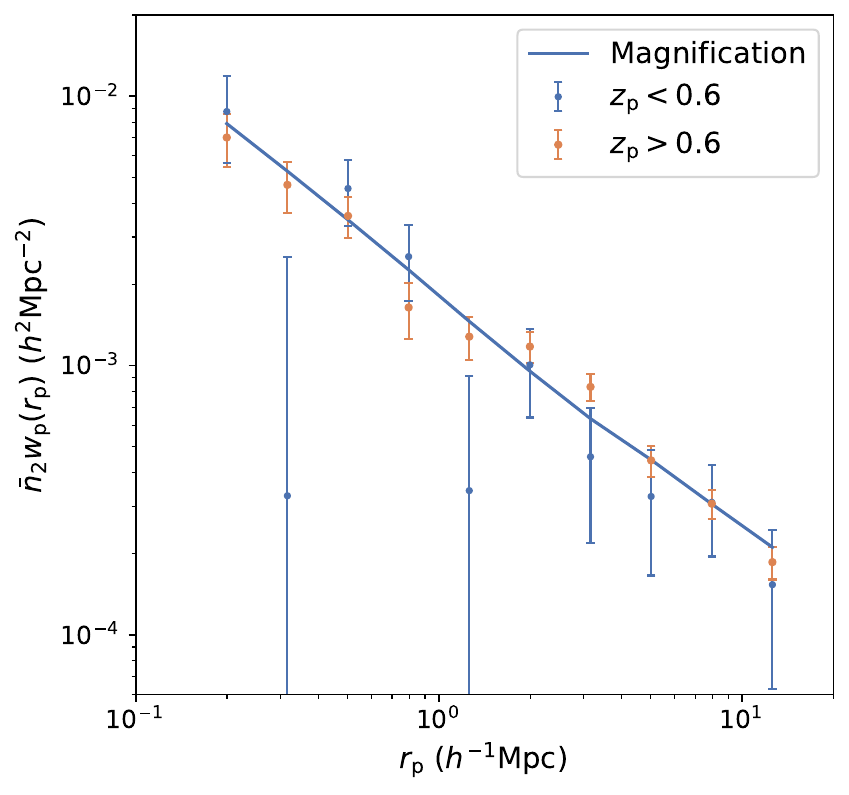}
    \caption{The measurement of $\bar{n}_{2}w_{\rm p}$ around quasars for foreground photometric galaxies with $z_{\rm p}<0.6$ in the stellar-mass bin $11.6<\log_{10}(M_{\ast}/M_{\odot})<11.8$, compared with the magnification prediction from the simulation.  In contrast, the signal from photometric galaxies with $z_{\rm p}>0.6$ is dominated by physical clustering.  For display, both the measured $z_{\rm p}<0.6$ signal and the magnification prediction are multiplied by $-1$.}
    \label{fig:0.0-0.6mag}
\end{figure}

\begin{figure}[!t]
    \centering
    \includegraphics[width=1.0\linewidth]{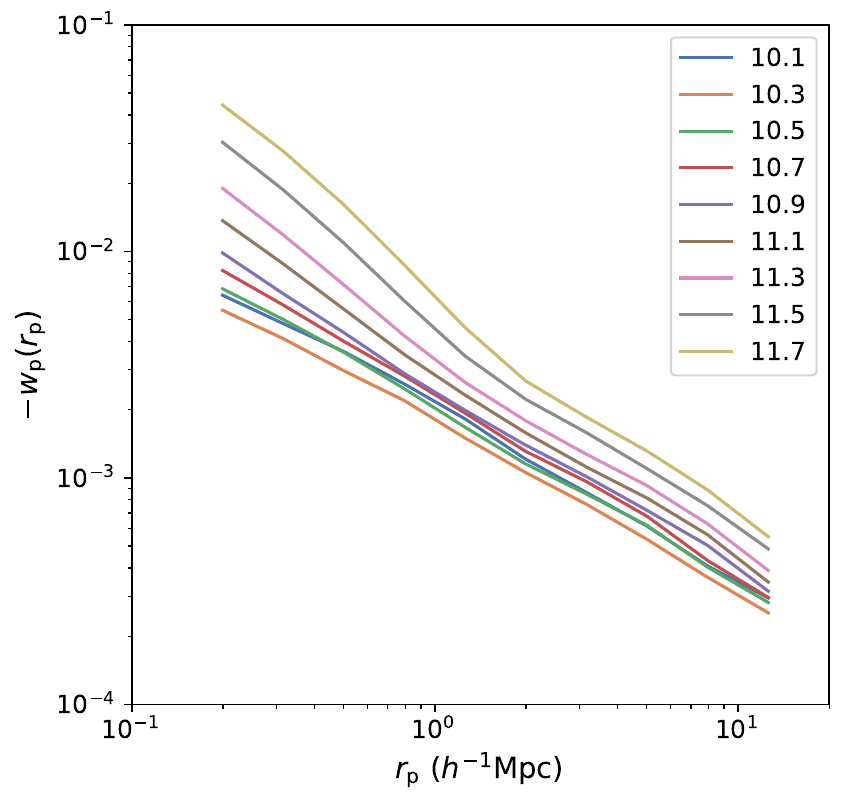}
    \caption{The measurement of $w_{\rm p}$ around quasars for intermediate-redshift photometric galaxies with $0.6<z_{\rm p}<0.8$, compared with the magnification prediction from the simulation.  This bin quantifies the residual foreground contribution that can remain after removing galaxies with $z_{\rm p}<0.6$.}
    \label{fig:0.6-0.8mag}
\end{figure}

As a diagnostic, we first measure $\bar{n}_{2}w_{\rm p}$ around quasars using foreground photometric galaxies with $z_{\rm p}<0.6$. Fig.~\ref{fig:0.0-0.6mag} shows this comparison for the highest stellar-mass bin, $11.6<\log_{10}(M_{\ast}/M_{\odot})<11.8$; for display, both the measured $z_{\rm p}<0.6$ signal and the simulated magnification prediction are multiplied by $-1$. Their agreement supports the interpretation that this low-redshift contribution is dominated by magnification. The signal from photometric galaxies with $z_{\rm p}>0.6$ is instead expected to be predominantly physical quasar--galaxy clustering, although a residual magnification term remains. We also estimate the contribution from galaxies with $0.6<z_{\rm p}<0.8$ in the same nine stellar-mass bins shown in Fig.~\ref{fig:0.6-0.8mag}. This residual foreground term is small compared with the physical quasar--galaxy signal on the scales used for the model fits.

\begin{figure*}[!t]
    \centering
    \includegraphics[width=0.8\linewidth]{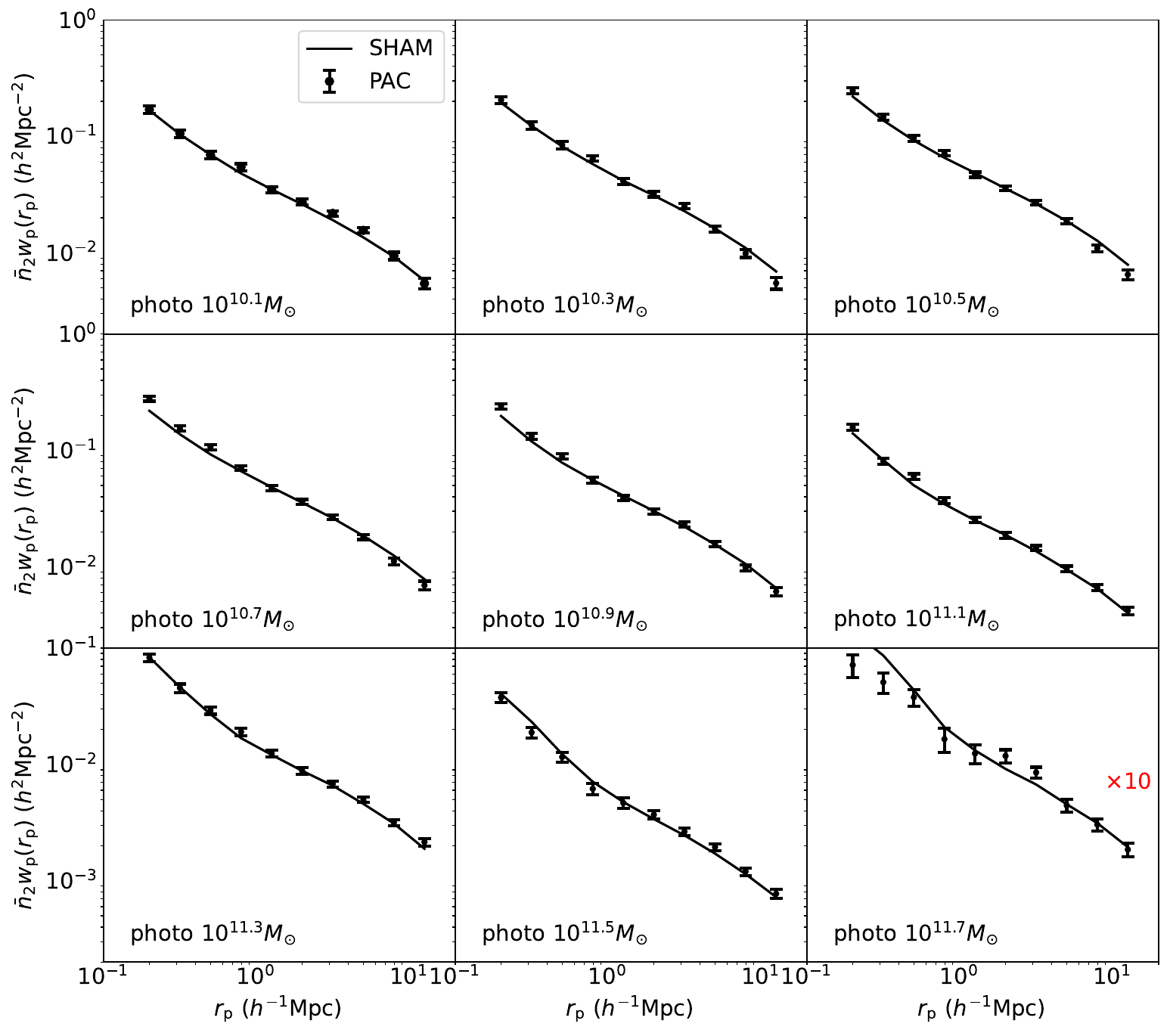}
    \caption{Measurements (points with error bars) and best-fitting model (lines) for $\bar{n}_{2}w_{\rm p}$ around DESI DR1 quasars after removing photometric galaxies with $z_{\rm p}<0.6$ to mitigate foreground magnification mostly. The galaxy stellar-mass range is $10^{10.0}\leq M_{\ast}/M_{\odot}\leq10^{11.8}$, with the median stellar mass indicated in each panel. The highest-stellar-mass bin is multiplied by 10 for visual clarity.}
    \label{fig:withmag_post}
\end{figure*}

We remove photometric galaxies with $z_{\rm p}<0.6$ from the PAC sample and repeat both the measurement and the MCMC modelling. This cut suppresses the dominant foreground-magnification contribution while retaining most galaxies that can be physically associated with the quasar redshift range. Some residual magnification from galaxies at $0.6<z_{\rm p}<0.8$ remains, but the simulation indicates that it is a small fraction of the total signal on the scales used in the fit. We present both the full-sample and magnification-mitigated fits in Table~\ref{tab:mcmc}; the latter measurement is shown in Fig.~\ref{fig:withmag_post}.

\begin{figure*}[!t]
    \centering
    \includegraphics[width=1.0\textwidth]{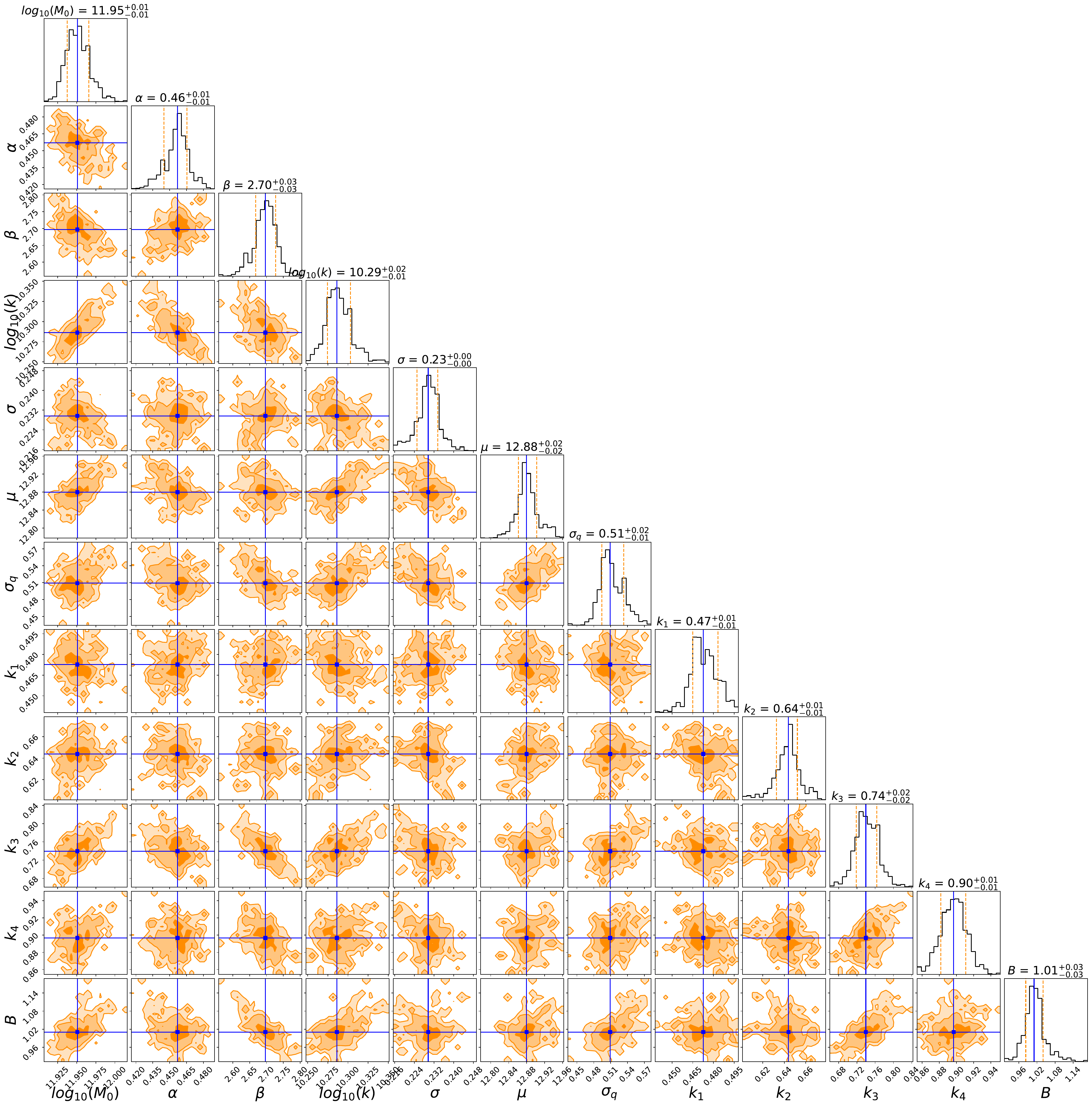}
    \caption{Marginalized posterior distributions (diagonal panels) and joint posterior contours (off-diagonal panels) for the fiducial magnification-mitigated fit using photometric galaxies with $z_{\rm p}>0.6$. Blue lines and markers indicate posterior medians, while orange dashed lines in the diagonal panels mark the 16th and 84th percentiles.}
    \label{fig:withmag_contour_post}
\end{figure*}

\subsection{SHMR and quasar--halo connection}\label{subsec:shmr_qhalo}

The constraints on the SHMR, incompleteness factors, and quasar--halo connection are summarized in Table~\ref{tab:mcmc}. The first row gives the fit to the full photometric sample, while the second gives the fit after removing galaxies with $z_{\rm p}<0.6$. The model is constrained jointly by the excess projected density $\bar{n}_{2}w_{\rm p}$, the projected autocorrelation functions, and the quasar--LRG projected cross-correlation. The best-fitting predictions are compared with the measurements in Figs.~\ref{fig:withmag_pre}--\ref{fig:cross_pre} and Fig.~\ref{fig:withmag_post}. Figure~\ref{fig:withmag_contour_post} shows the marginalized posterior distributions and joint parameter constraints for the fiducial magnification-mitigated fit, illustrating the covariances among the SHMR, incompleteness, and quasar--halo parameters.

For the fit to the full photometric sample, we obtain $\log_{10}(M_0/h^{-1}M_{\odot})=11.97^{+0.02}_{-0.02}$, with SHMR slopes $\alpha=0.41^{+0.02}_{-0.01}$ and $\beta=2.72^{+0.04}_{-0.03}$, normalization $\log_{10}(k/M_{\odot})=10.26^{+0.02}_{-0.03}$, and scatter $\sigma=0.26^{+0.01}_{-0.01}$. The inferred quasar-hosting probability peaks at $\mu=12.76^{+0.03}_{-0.04}$ with width $\sigma_{\rm q}=0.48^{+0.04}_{-0.03}$, where both quantities are defined in $\log_{10}(M_{\rm acc}/h^{-1}M_{\odot})$. The relative satellite-hosting parameter is $B=1.19^{+0.07}_{-0.06}$.

For the magnification-mitigated fit, the SHMR parameters shift mildly to $\log_{10}(M_0/h^{-1}M_{\odot})=11.95^{+0.01}_{-0.01}$, $\alpha=0.46^{+0.01}_{-0.01}$, $\beta=2.70^{+0.03}_{-0.03}$, $\log_{10}(k/M_{\odot})=10.29^{+0.02}_{-0.01}$, and $\sigma=0.23^{+0.00}_{-0.00}$. The corresponding quasar--halo parameters are $\mu=12.88^{+0.02}_{-0.02}$, $\sigma_{\rm q}=0.51^{+0.02}_{-0.01}$, and $B=1.01^{+0.03}_{-0.03}$.

The incompleteness factors for the four lowest-stellar-mass bins are also constrained. For the full sample, we find $k_1=0.43^{+0.02}_{-0.02}$, $k_2=0.61^{+0.02}_{-0.03}$, $k_3=0.67^{+0.03}_{-0.02}$, and $k_4=0.87^{+0.02}_{-0.03}$. For the magnification-mitigated measurement, these values become $k_1=0.47^{+0.01}_{-0.01}$, $k_2=0.64^{+0.01}_{-0.01}$, $k_3=0.74^{+0.02}_{-0.02}$, and $k_4=0.90^{+0.01}_{-0.01}$. These factors allow the lower-mass photometric sample to contribute to the SHMR constraints without treating it as mass-complete.

Overall, the inferred SHMR is comparatively stable against the foreground-magnification treatment, whereas the quasar--halo parameters are more sensitive to it. The magnification-mitigated value $B\simeq1$ indicates no statistically required enhancement or suppression of quasar activity in satellite subhalos relative to central halos at fixed halo accretion mass. The combined PAC measurements therefore constrain the galaxy--halo connection, stellar-mass incompleteness, and quasar occupation simultaneously at $0.8<z_{\rm s}<1.0$.

\begin{table*}
\caption{Posterior median and 16th--84th percentile constraints on the SHMR, stellar-mass incompleteness, and quasar--halo connection for a Gaussian quasar occupation model.\label{tab:mcmc}}
\centering
\resizebox{\textwidth}{!}{%
\begin{tabular}{l*{12}{c}}
\hline\hline
Photometric sample & $\log_{10}(M_0/h^{-1}M_{\odot})$ & $\alpha$ & $\beta$ & $\log_{10}(k/M_{\odot})$ & $\sigma$ & $\mu$ & $\sigma_{\rm q}$ & $k_1$ & $k_2$ & $k_3$ & $k_4$ & $B$ \\
\hline
Full sample & $11.97_{-0.02}^{+0.02}$ & $0.41_{-0.01}^{+0.02}$ & $2.72_{-0.03}^{+0.04}$ & $10.26_{-0.03}^{+0.02}$ & $0.26_{-0.01}^{+0.01}$ & $12.76_{-0.04}^{+0.03}$ & $0.48_{-0.03}^{+0.04}$ & $0.43_{-0.02}^{+0.02}$ & $0.61_{-0.03}^{+0.02}$ & $0.67_{-0.02}^{+0.03}$ & $0.87_{-0.03}^{+0.02}$ & $1.19_{-0.06}^{+0.07}$ \\
$z_{\rm p}>0.6$ & $11.95_{-0.01}^{+0.01}$ & $0.46_{-0.01}^{+0.01}$ & $2.70_{-0.03}^{+0.03}$ & $10.29_{-0.01}^{+0.02}$ & $0.23_{-0.00}^{+0.00}$ & $12.88_{-0.02}^{+0.02}$ & $0.51_{-0.01}^{+0.02}$ & $0.47_{-0.01}^{+0.01}$ & $0.64_{-0.01}^{+0.01}$ & $0.74_{-0.02}^{+0.02}$ & $0.90_{-0.01}^{+0.01}$ & $1.01_{-0.03}^{+0.03}$ \\
\hline
\end{tabular}}
\tablefoot{The second row excludes photometric galaxies with $z_{\rm p}<0.6$ to mitigate foreground magnification. The unit of $M_0$ is $h^{-1}M_{\odot}$ and that of $k$ is $M_{\odot}$.}
\end{table*}

\section{Discussion}\label{sec:discussion}

The principal quasar--halo result is $B=1.01^{+0.03}_{-0.03}$, complementing recent quasar halo-occupation analyses based on DESI clustering and cosmological simulations \citep{yuan2024desihod}. In the model of Equation~\ref{eq:fsat}, $B$ is not itself the quasar satellite fraction, which also depends on the relative abundances of selected central halos and subhalos in the simulation. Instead, $B$ measures the relative quasar-hosting probability of satellite subhalos and central halos at fixed halo accretion mass.

At $z\simeq0.9$, this result does not require an additional central--satellite dependence of quasar occupation once halo accretion mass is matched \citep{chowdhary2025qsohod}. Previous quasar halo-occupation studies have generally found that the satellite component is difficult to constrain with spectroscopic quasar pairs alone and that a wide range of satellite fractions can be allowed depending on the assumed occupation model \citep{richardson2012,chatterjee2013,shen2013,eftekharzadeh2019}. PAC provides a complementary route to this question: the dense photometric galaxy sample enables precise measurements of quasar environments on sub-Mpc scales.

The preferred occupation peak, $\mu=12.88^{+0.02}_{-0.02}$, lies at a few $10^{12}\,h^{-1}M_{\odot}$ and is broadly consistent with large-scale quasar clustering measurements that place quasars in moderately massive halos \citep{croom2005,ross2009,shen2009,white2012,eftekharzadeh2015,laurent2017,hou2021}. This parameter describes the peak of the assumed occupation function rather than a directly measured mean host-halo mass. Halos at this scale are relevant to models of SMBH--galaxy coevolution, but our clustering analysis does not by itself determine feedback modes or outflow energetics; it constrains the halo environments in which luminous SMBH accretion occurs \citep{kormendy2013,heckman2014,fabian2012,king2015,laha2021}.

The comparison between the full-sample and magnification-mitigated fits illustrates why foreground magnification must be treated carefully. Including all foreground photometric galaxies gives $B=1.19^{+0.07}_{-0.06}$, whereas removing galaxies with $z_{\rm p}<0.6$ gives $B=1.01^{+0.03}_{-0.03}$. Because magnification adds a quasar--galaxy correlation that is not physical clustering, it can bias the inferred one-halo occupation if absorbed by the quasar--halo model. The foreground-galaxy diagnostic in Fig.~\ref{fig:0.0-0.6mag} supports the adopted interpretation and motivates the magnification-mitigated measurement as our fiducial constraint. The interpretation of these results remains conditional on several modelling and observational assumptions. We use a Gaussian occupation probability in $\log M_{\rm acc}$ and a single relative satellite factor $B$; the model does not capture possible dependencies on quasar luminosity, colour, accretion rate, obscuration, host-galaxy star-formation state, or halo assembly history.

\section{Conclusions}\label{sec:conclusions}

We applied the PAC method to DESI DR1 quasars at $0.8<z_{\rm s}<1.0$ and photometric galaxies from the DESI Legacy Imaging Surveys DR9. Exploiting the dense photometric sample, PAC measures the excess projected surface density of galaxies around quasars over $0.1<r_{\rm p}/(h^{-1}\,\mathrm{Mpc})<15$ down to $M_{\ast}=10^{10.80}M_{\odot}$. We jointly interpret these measurements with the quasar and LRG autocorrelation functions and the quasar--LRG cross-correlation using an N-body simulation, a stellar-to-halo mass relation, stellar-mass-incompleteness parameters, and a Gaussian quasar occupation model. Our main results are as follows.
\begin{enumerate}
    \item The Gaussian quasar occupation peaks at $\log_{10}(M_{\rm acc}/h^{-1}M_{\odot})=12.88^{+0.02}_{-0.02}$ with width $\sigma_{\rm q}=0.51^{+0.02}_{-0.01}$. The relative satellite-hosting parameter, defined as the quasar-hosting probability of a satellite subhalo relative to that of a central halo at fixed halo accretion mass, is $B=1.01^{+0.03}_{-0.03}$.

    \item The SHMR is constrained by $\log_{10}(M_0/h^{-1}M_{\odot})=11.95^{+0.01}_{-0.01}$, $\alpha=0.46^{+0.01}_{-0.01}$, $\beta=2.70^{+0.03}_{-0.03}$, $\log_{10}(k/M_{\odot})=10.29^{+0.02}_{-0.01}$, and intrinsic scatter $\sigma=0.23^{+0.00}_{-0.00}$.

    \item Within the adopted model framework, quasars are consistent with being equally likely to reside in central halos and satellite subhalos at fixed halo accretion mass.
\end{enumerate}

These results demonstrate the potential of PAC for precise small-scale measurements of quasar environments. Future analyses using the full five-year DESI data set will reduce statistical uncertainties and enable tests of whether quasar halo occupation depends on luminosity, colour, emission-line properties, or accretion state.

\begin{acknowledgements}
S.G. thanks Yipeng Jing, Simon D. M. White, Hongyu Gao, and Yun Zheng for valuable discussions and comments.

This work was supported by the China Scholarship Council–German Academic Exchange Service (CSC–DAAD) Joint Scholarship Program, the National Natural Science Foundation of China (NSFC; Grants 12133006 and 11890691), the National Key Research and Development Program of China (Grants 2023YFA1607800 and 2023YFA1607801), Grant No. CMS-CSST-2021-A03, and the 111 Project (Grant No. B20019). The computations in this work were carried out on the Gravity Supercomputer at the Department of Astronomy, Shanghai Jiao Tong University, and on computing resources provided by the Max Planck Computing and Data Facility (MPCDF) through the Max Planck Institute for Astrophysics (MPA).

This research is based on data obtained with the Dark Energy Spectroscopic Instrument (DESI). DESI is constructed and operated by the DESI Collaboration. Funding for the DESI Project has been provided by the U.S. Department of Energy, Office of Science, Office of High Energy Physics; the National Energy Research Scientific Computing Center (NERSC), a U.S. Department of Energy Office of Science User Facility; the U.S. National Science Foundation, Division of Astronomical Sciences; the Science and Technology Facilities Council of the United Kingdom; the Gordon and Betty Moore Foundation; the Heising-Simons Foundation; the French Alternative Energies and Atomic Energy Commission (CEA); the National Council of Humanities, Science and Technology of Mexico (CONAHCYT); the Ministry of Science and Innovation of Spain; and the DESI Member Institutions.

The DESI Legacy Imaging Surveys consist of three complementary projects: the Dark Energy Camera Legacy Survey (DECaLS; Proposal ID 2014B-0404; PIs: David Schlegel and Arjun Dey), the Beijing--Arizona Sky Survey (BASS; NOAO Proposal ID 2015A-0801; PIs: Zhou Xu and Xiaohui Fan), and the Mayall $z$-band Legacy Survey (MzLS; Proposal ID 2016A-0453; PI: Arjun Dey).
\end{acknowledgements}

\bibliographystyle{aa}
\bibliography{references}

\end{document}